\UseRawInputEncoding
\documentclass[journal=jpcafh,manuscript=article]{achemso}

\usepackage{graphicx}
\usepackage{subfig} 
\usepackage{epstopdf} 

\usepackage{amssymb}
\usepackage{amsmath}

\let\oldAA\AA
\renewcommand{\AA}{\text{\normalfont\oldAA}}

\usepackage{threeparttable,array,booktabs,calc}
\usepackage{dcolumn}
\usepackage{bm}

\usepackage[version=4]{mhchem}

\usepackage{hyperref}

\author{Katrin Dulitz}
\affiliation{Institute of Physics, University of Freiburg, Hermann-Herder-Str. 3, 79104 Freiburg, Germany}
\email{katrin.dulitz@physik.uni-freiburg.de}
\author{Marco van den Beld-Serrano}
\affiliation{Institute of Physics, University of Freiburg, Hermann-Herder-Str. 3, 79104 Freiburg, Germany}
\author{Frank Stienkemeier}
\affiliation{Institute of Physics, University of Freiburg, Hermann-Herder-Str. 3, 79104 Freiburg, Germany}

\title[]{Single-Source, Collinear Merged-Beam Experiment for the Study of Reactive Neutral-Neutral Collisions}

\abbreviations{}

\begin{document}

%
%
%
%
%

\begin{abstract}
We present two methods for studying reactive collisions between two atomic or molecular species: a collinear merged-beam method, in which two gas pulses from a single supersonic beam source are coalesced, and an intrabeam-scattering technique, in which a single gas pulse is used. Both approaches, which rely on the laser cooling and deceleration of a laser-coolable species inside a Zeeman slower, can be used for a wide range of scattering studies. Possible experimental implementations of the proposed methods are outlined for autoionizing collisions between helium atoms in the metastable $2^3\mathrm{S}_1$ state and a second, atomic or molecular species. Using numerical trajectory calculations, we provide estimates of the expected on-axis detection efficiency, collision-energy range and collision-energy resolution of the approach. We have experimentally tested the feasibility of such an experiment by producing two gas pulses at very short time intervals, and the results of these measurements are detailed as well.
\end{abstract}

\section{\label{sec:intrometastables}Introduction}
The study of atomic and molecular collisions in the few-partial wave regime offers intriguing insights into the physical nature of a collision complex. At collision energies $E_\mathrm{c}/k_\mathrm{B}$ (where $k_\mathrm{B}$ denotes the Boltzmann constant) below 1 K, purely quantum-mechanical phenomena -- such as barrier tunneling and quantum reflection -- can be observed, since the long-range centrifugal barrier becomes very shallow and individual partial-wave contributions are no longer washed out due to thermal averaging \cite{Naulin2014, Wang2018}. The low-energy conditions also allow for the formation of exotic long-range molecules which can potentially be studied in supersonic beam experiments using photoassociation spectroscopy \cite{Skomorowski2016, Jones2006}.

Gas-phase collision studies have traditionally been carried out using crossed atomic or molecular beams \cite{Herschbach1987}. For such an experiment, the collision energy in the center-of-mass frame of reference is given by
\begin{equation}
	E_{\mathrm{c}} = \frac{\mu}{2}\left(   |\vec{v}_1|^2 + |\vec{v}_2|^2 - 2|\vec{v}_1||\vec{v}_2|\cos(\alpha)   \right),
	\label{eq:Ecoll}
\end{equation}
where $\mu$ is the reduced mass and $\alpha$ denotes the angle between the velocity vectors $\vec{v}_1$ and $\vec{v}_2$ of the two beams. In this article, we frequently use the relative velocity $v_{\mathrm{rel}} = \sqrt{2E_{\mathrm{c}}/\mu}$ instead of the collision energy, as it is independent of the reduced mass.

From Eq. \ref{eq:Ecoll}, we can see that the collision energy can be tuned either by changing the beam velocities or by varying the collision angle. The production of low-velocity supersonic beams for the study of cold neutral-neutral collisions is not straightforward. The CRESU technique \footnote{cin\'{e}tic de r\'{e}action en \'{e}coulement supersonique uniforme, or reaction kinetics in uniform supersonic flow}, which relies on an isentropic gas expansion through a Laval nozzle, has been developed in the 1980s for studying the dynamics of astrochemically relevant ion-molecule and neutral-neutral reactions \cite{Sims1995, Potapov2017}. In recent years, several techniques have been put forward to decelerate atomic and molecular beams, such as Stark and Zeeman deceleration, counter-rotating nozzles, the photostop technique or kinematic slowing \cite{Bell2009, Schnell2009, Hogan2011, van2012}. However, for the study of collisions in the quantum regime, the velocities of both beams have to be reduced to near zero, which results in a low flux of decelerated particles in both beams \cite{van2009}. However, a number of collision-energy-resolved inelastic collision experiments have already been realized by intersecting a Stark-decelerated molecular beam with a supersonic beam emerging from a pulsed valve \cite{van2012, Vogels2015a, Vogels2018}. In the second approach for studying low-energy collisions, the relative angle $\alpha$ is set to a value $<\,90^{\circ}$. 
In the limit of parallel beams, the lowest collision energies can be reached, and Eq. \ref{eq:Ecoll} reduces to
\begin{equation}
	E_{\mathrm{c}} = \frac{\mu}{2}\left(|\vec{v}_1|-|\vec{v}_2|\right)^2.
	\label{eq:Ecollparallel}
\end{equation}
The relative angle vanishes when two supersonic beams are overlapped in a merged-beam setup \cite{Phaneuf1999}. This method, which has first been applied to ion-molecule collisions \cite{Trujillo1966, Neynaber1968, Belyaev1966}, has recently been extended to the study of quantum resonances and collision stereodynamics in autoionizing collisions between long-lived, excited state (metastable) noble gases and different atomic or molecular species \cite{Henson2012, Jankunas2015b, Gordon2018}. In these experiments, the trajectories of two supersonic beams are overlapped using electromagnetic guiding fields. A different merged-beam approach, which relies on the use of a counter-rotating nozzle in combination with a second beam source, has been proposed by Herschbach and co-workers \cite{Sheffield2012, Wei2012}. Besides that, it has also been possible to study low-energy ion-molecule collisions in a merged-beam setup using the Rydberg-deceleration technique \cite{Allmendinger2016}. Merged-beam experiments were also performed in the group of H. L. Bethlem using a storage ring that was tangentially crossed by a second supersonic beam \cite{vanderPoel2018}.

Cold and ultracold collisions have also been observed inside a single supersonic beam. Such ``intrabeam collisions'' occur in the presence of different velocity classes, e.g. while the beam is actively slowed by laser cooling \cite{Weiner1999, TaillandierLoize2018} or owing to the velocity slip effect which is caused by the co-expansion of two atomic or molecular species of different mass \cite{Amarasinghe2017, Perreault2017, Perreault2018b}.

Here, we propose two methods for studying thermal and cold reactive collisions. The first method combines the advantages of merged-beam and intrabeam collision experiments. Rather than achieving beam overlap by transversely merging two beams from different sources using electromagnetic fields, this approach relies on the longitudinal merging of two atomic or molecular beams produced by two subsequent gas pulses of the same supersonic jet source. In order to merge the two gas pulses, the velocity of the atoms in the first pulse -- which is He($2^3\mathrm{S}_1$) in our specific case -- is decreased by laser cooling and deceleration in a Zeeman slower. The collision energy is tuned by simultaneously changing the relative time delay between the gas pulses and the duration of the Zeeman slowing process. Using the second method, low-energy (intrabeam) collisions between two species can be studied at tunable energy inside a single gas pulse by Zeeman slowing one of the species contained in the beam.

In this article, the methods are outlined for autoionizing collisions between metastable He atoms in the $2^3\mathrm{S}_1$ state and a second, atomic or molecular species. However, the proposed techniques are, in principle, also suitable for collision studies involving other laser-coolable species (see conclusion section). During an autoionizing collision, ionization is a result of collisional energy transfer between an electronically excited species, typically a metastable rare gas atom (Rg$^*$), and an atom or molecule (M). Among the various reaction channels, Penning ionization and associative ionization are the most common ones at thermal and at low energies:
\begin{equation}
	\mathrm{Rg^*+M \rightarrow [RgM]^*\rightarrow}
	\left\{  
	\begin{matrix}
		\mathrm{Rg + M^+ + e^-}\,\,\mathrm{(Penning\,\,ionization)}\\
		\mathrm{RgM^+ + e^-}\,\,\mathrm{(Associative\,\,ionization)}
	\end{matrix}
	\right.
\end{equation}
Collisions of He($2^3\mathrm{S}_1$) are of particular scientific interest, since He($2^3\mathrm{S}_1$) can ionize any atom or molecule except for Ne owing to its high electronic energy of 19.8 eV \cite{NIST_ASD}. In addition to that, the laser cooling of He($2^3\mathrm{S}_1$) via the $2^3\mathrm{P}_2 \leftarrow 2^3\mathrm{S}_1$ transition has been demonstrated in a large number of experiments \cite{Vassen2012}; and even the Bose-Einstein condensation of this species has been achieved \cite{Robert2001,PereiraDosSantos2001}.

We describe possible experimental implementations of the proposed methods. We use numerical trajectory calculations to provide estimates of the expected on-axis detection efficiency, velocity-tuning range and collision velocity resolution of the approach at thermal collision energies and at low collision energy. We also show the results of experimental measurements, in which we have experimentally tested the feasibility of producing two gas pulses at very short time intervals ($\leq$ 2000 $\mu$s).

\section{\label{sec:generaloutline}Experiment}
\begin{figure*}[!ht] 
	\centering
	\includegraphics{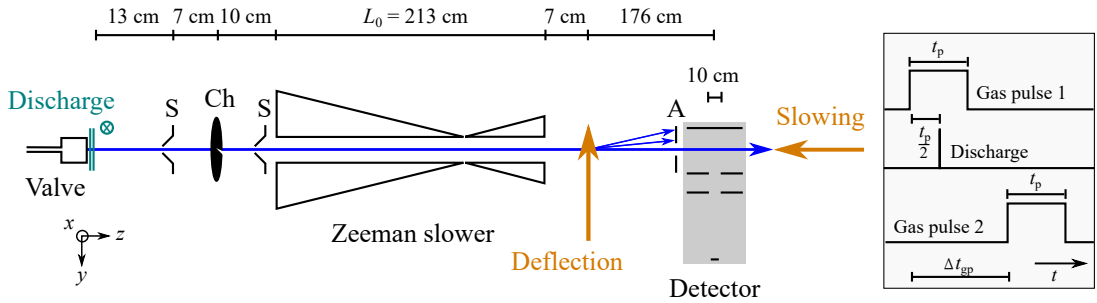}
	\caption[]{Schematic drawing of the proposed experimental setup, not to scale. Inset: time sequence for the operation of the pulsed valve and the discharge. S = skimmer, Ch = chopper wheel, A = aperture.}
	\label{fig:propsetup}
\end{figure*}

A sketch of the proposed experimental setup for merged-beam collisions is shown in Fig. \ref{fig:propsetup}. Supersonic jets at a mean velocity $|\vec{v}_0| = v_0 = 1000$ m/s and at a longitudinal velocity spread of $\Delta v_{\mathrm{FWHM}}$ = 107 m/s full width at half maximum are generated by co-expanding an atomic or molecular species in He gas from a cryogenically cooled pulsed valve (e.g., a CRUCS valve \cite{Grzesiak2018}) maintained at high backing pressure. The specific valve characteristics have been determined in numerous experiments in our laboratory \cite{Grzesiak2018, Grzesiak2019}. Here, we assume a typical gas pulse duration of $t_\mathrm{p} = 20\,\mu$s \footnote{In the experiment, the pulse duration $t_\mathrm{p}$ is obtained by measuring the voltage drop on a probe head during discharge excitation.}. To achieve $v_0 = 1000$ m/s, which is a typical initial velocity used for the laser cooling of He($2^3$S$_1$) \cite{Rooijakkers1996}, the valve needs to be cooled to cryogenic temperatures. However, the use of higher initial velocities is unfeasible, as the stopping distance for Zeeman slowing rapidly increases as the initial velocity is increased, e.g., from $L_0$ = 213 cm at $v_0 = 1000$ m/s to $L_0$ = 479 cm at $v_0 = 1500$ m/s.

Two gas pulses are generated at two time instants $t_1$ and $t_2$ with $\Delta t_{\mathrm{gp}} = t_2-t_1$. He atoms in the metastable $2^3\mathrm{S}_1$ and $2^1\mathrm{S}_0$ states are produced by a pulsed electron-seeded gas discharge in the central part of pulse 1 (at $t_\mathrm{p}/2$, where $t_\mathrm{p}$ is the duration of the gas pulse). Since the discharge pulse duration can be made as short as $1\,\mu$s \cite{Lewandowski2004}, we assume an infinitesimally short discharge pulse duration for simplicity. Beam 2 is not discharge-excited and thus contains the atomic or molecular species and ground-state He atoms only.

In order to merge the two gas pulses, the mean velocity of the He($2^3\mathrm{S}_1$) atoms in pulse 1 ($|\vec{v}_1| = v_1$) is decreased by laser cooling and deceleration in a Zeeman slower using a counter-propagating laser beam resonant with the $2^3$S$_1$ -- $2^3$P$_2$ transition in He at 1083 nm. To control the amount of deceleration and thus the collision energy, the laser light is only switched on for a certain time period $\Delta t_{\mathrm{Zee}}$, e.g., by switching the radiofrequency power to an acousto-optical modulator between high and low. Since He atoms in the $2^1$S$_0$ state and other co-expanding atomic or molecular species contained in pulse 1 are not addressed by the laser radiation, they are separated from He atoms in the $2^3$S$_1$ state by their difference in time of flight to the detector. Therefore, this method is intrinsically quantum-state- and species-selective. Since pulse 2 does not contain He($2^3\mathrm{S}_1$), its mean velocity remains unchanged at all times, i.e., $|\vec{v}_2| = v_2 = v_0$. 

The reaction products (ions, electrons) are accelerated onto a detector (active diameter of 10 cm) positioned in a direction perpendicular to the supersonic beam axis. In this way, only collisions inside the detection region (4.21 $\leq z \leq$ 4.31 m) are monitored. To increase the product yield, it is also possible to place a sequence of detectors along the beam axis or to use electric focusing fields to sample a larger interaction volume. An in-line detector may not be suitable, since metastable atoms cause a large secondary-electron signal when they impinge on the metal surface of a detector. 

Since Zeeman slowing mainly affects the longitudinal velocity component of the supersonic beam, the transverse divergence of the beam containing He($2^3$S$_1$) increases as the longitudinal velocity of the atoms is decreased. To ensure a good spatial overlap of pulse 1 and pulse 2 in the detection zone, the transverse velocity of the beam containing He($2^3$S$_1$) can additionally be reduced using optical collimation in a zig-zag or in a curved wavefront configuration in front of the skimmer and behind the Zeeman slower (see e.g. \cite{Aspect1990, Rooijakkers1996}). In the simulations described below, we assume one-dimensional velocity distributions along the supersonic beam axis for simplicity.

To further improve the collision energy resolution, He($2^3$S$_1$) atoms, whose longitudinal velocities are outside the desired range for collision studies, can be deflected by laser radiation resonant with the $2^3$S$_1$ -- $2^3$P$_2$ transition. In addition to that, the trailing edges of pulse 2 can be cut off using a rapidly rotating chopper wheel. Using a chopper wheel as described in Ref. \cite{Lam2015} (maximum rotation speed $S_{\mathrm{max}}$ = 1.3 kHz, 1 mm slit width), the gas pulse duration can be reduced to a minimum of 13 $\mu$s. The position of the chopper wheel along the beam axis is chosen such that all metastable He atoms inside of pulse 1 pass through the chopper wheel, while the pulse duration of pulse 2 is significantly reduced.

\section{Results}
\subsection{\label{sec:velocitytuning}Collision-energy tuning}
For all simulations, we assume a constant initial beam velocity ($v_0 = 1000$ m/s) and fixed distances between the detector and the other parts of the setup (Fig. \ref{fig:propsetup}). In this case, the resulting final velocity of the He($2^3$S$_1$) atoms in pulse 1 (and thus the collision energy range) is determined by the time period, during which the Zeeman slower is turned on, $\Delta t_{\mathrm{Zee}}$. In addition to that, also the initial velocity $v_0$ could be varied. However, a detailed discussion of this additional parameter would go beyond the scope of the article. Apart from $\Delta t_{\mathrm{Zee}}$, the time difference between the two gas pulses $\Delta t_{\mathrm{gp}}$ must be adjusted to ensure that the collisional interaction takes place at the detector position. According to Eq. \ref{eq:Ecollparallel}, the lowest collision energies $E_{\mathrm{c}}\,\longrightarrow\,0$ are achieved when $v_1\,\longrightarrow\,v_2$, i.e., when $\Delta t_{\mathrm{Zee}}\,\longrightarrow\,0$ and $\Delta t_{\mathrm{gp}}\,\longrightarrow\,0$. Likewise, the maximum collision energy is reached when the  He($2^3$S$_1$) beam is decelerated to a near standstill ($v_1\,\longrightarrow\,0$). In practice, these limits cannot be reached due to technical constraints and due to the transverse divergence of the He($2^3$S$_1$) beam at very low velocities. Assuming that the maximum light scattering force is exerted on the particles during Zeeman slowing, the position-dependent velocity of pulse 1 is given by
\begin{equation}
	v_1(z) = v_0 \sqrt{1 - \frac{\left(z-z_{\mathrm{Zee}}\right)}{L_0}},
	\label{eq:v1}
\end{equation}
where $z_{\mathrm{Zee}}$ = 30 cm is the entrance position of the Zeeman slower and $L_0$ = 213 cm is the stopping distance in our proposed scheme. Here, we follow the explanations in Refs. \cite{Hopkins2016, BenAli2017} for the calculation of the radiation pressure inside a Zeeman slower.

\begin{figure*}[!ht] 
	\centering
	\includegraphics{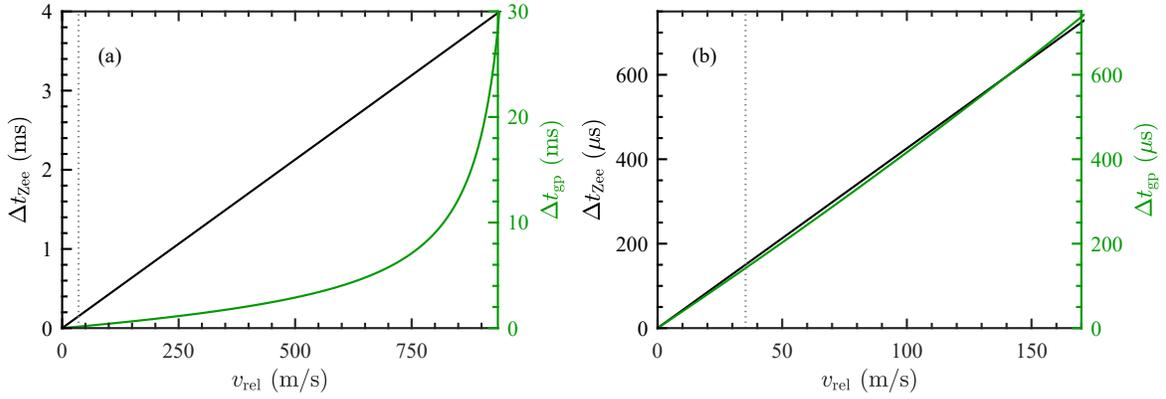}
	\caption[]{Time period for Zeeman slowing $\Delta t_{\mathrm{Zee}}$ (black color) and time difference between the two gas pulses $\Delta t_{\mathrm{gp}}$ (green color) as a function of relative beam velocity $v_{\mathrm{rel}}$ for collisions of an ideal He($2^3$S$_1$) atom of pulse 1 with an ideal particle of pulse 2 (a) over nearly the full collision energy range and (b) in a close-up view of the low-collision energy range. The gray dotted line is a vertical line at a relative velocity of 35 m/s.}
	\label{fig:vrel}
\end{figure*}

The time delay between the two gas pulses can then be obtained as follows:
\begin{equation}
	\Delta t_{\mathrm{gp}} = t_1 - t_2 = \left( t_\mathrm{s} + \Delta t_{\mathrm{Zee}} + \frac{z_\mathrm{c}-z_\mathrm{off}}{v_0-v_\mathrm{rel}} \right) - \left( \frac{z_\mathrm{c}}{v_0} \right),
	\label{eq:tgp}
\end{equation}
where $t_\mathrm{s}$ is the time of flight until the entrance of the Zeeman slower, $z_\mathrm{c}$ is the collision position and $z_\mathrm{off}$ is the particle position at which the Zeeman slower is turned off. By re-arranging Eq. \ref{eq:tgp}, the collision position is then obtained as
\begin{equation}
	z_\mathrm{c} = \frac{(v_0-v_\mathrm{rel})v_0}{v_\mathrm{rel}}\cdot \left( \Delta t_{\mathrm{gp}} + \frac{z_\mathrm{off}}{v_0-v_\mathrm{rel}} - \Delta t_{\mathrm{Zee}} \right)
	\label{eq:zcoll}
\end{equation}
The values of $\Delta t_{\mathrm{Zee}}$ and $\Delta t_{\mathrm{gp}}$ at different relative beam velocities $v_{\mathrm{rel}}$ are obtained from a numerical trajectory calculation for an ideal He($2^3$S$_1$) atom ($v_0$ = 1000 m/s, $t_0 = t_\mathrm{p}/2$) which collides with a second ideal particle of pulse 2 ($v_0$ = 1000 m/s, $t_0 = t_\mathrm{p}/2 + \Delta t_{\mathrm{gp}}$) in the center of the detector ($z_\mathrm{c}$ = 4.26 m). From Fig. \ref{fig:vrel} (a), we can see that $0\,\leq\,\Delta t_{\mathrm{gp}}\,\leq\,30$ ms and $0\,\leq\,\Delta t_{\mathrm{Zee}}\,\leq\,4$ ms are required to cover nearly the entire possible range of collision energies. In this article, the relative beam velocity $v_{\mathrm{rel}}$ is used instead of $E_{\mathrm{c}}$, since it is independent of the reduced mass of the particles. To give a reference value, $v_{\mathrm{rel}}$ must be $\leq$ 110 m/s to reach the cold collision regime ($T_\mathrm{coll}\,\leq\,1$ K) for He($2^3$S$_1$)-H$_2$ collisions.
In principle, it is also possible to further extend the distance between the Zeeman slower and the detection region, so that a lower collision energy is obtained for the same value of $\Delta t_{\mathrm{gp}}$. However, the overall length of the proposed setup is already comparable to the size of a typical laboratory.

For a more realistic estimate of the range of achievable relative velocities, the collision velocity resolution $\sigma_{v_{\mathrm{rel}}}/v_{\mathrm{rel}}$ and the detection efficiency $\vartheta$, we have to take into account the initial velocity and the time distribution of the particles. For this reason, we have calculated the distribution of collision positions and relative velocities along the beam axis using Monte-Carlo-type trajectory simulations of point-like particles (in MATLAB language). For each supersonic beam, all particles are assumed to originate from the same initial Gaussian velocity distribution produced at the same on-axis position $z_0\,=\,0$. The mean value $v_0 = 1000$ m/s and the standard deviation of each initial velocity distribution $\sigma_0 = \Delta v_{\mathrm{FWHM}}/(2\sqrt{2 \ln 2}) = 45$ m/s are experimentally measured values from our laboratory. All particles of pulse 1 are assumed to be produced at the same time $t_0 = t_\mathrm{p}/2$; particles of pulse 2 are uniformly generated during the time interval $\Delta t_{\mathrm{gp}} + \left(0 \leq t_0 \leq t_\mathrm{p}\right)$ (see Fig. \ref{fig:propsetup}). The values of $\Delta t_{\mathrm{Zee}}$ and $\Delta t_{\mathrm{gp}}$ are obtained for collisions between two ideal particles, as described above. The values of $\Delta t_{\mathrm{gp}}$ are slightly offset (by $\leq\pm 2$ \% of $\Delta t_{\mathrm{gp}}$) to ensure that the collision distribution is centered at the detector position. For simplicity, each particle of pulse 1 is assumed to collide with each particle of pulse 2. Therefore, the total number of collisions along the beam axis is $N_{\mathrm{c,tot}} = N_{1,0} \cdot N_{2,0}$, where $N_{1,0}$ ($N_{2,0}$) is the initial number of particles in pulse 1 (pulse 2). The detection efficiency is then given by $\vartheta = N_{\mathrm{c,d}}/N_{\mathrm{c,tot}}$, where $N_{\mathrm{c,d}}$ is the number of collisions in the detection region. The detection efficiency for the product ions is assumed to be unity.
Here, we also do not consider the transverse velocity distributions of the two supersonic beams. In the experiment, the transverse extent of both supersonic beams in the interaction zone can be made small using optical collimation (for pulse 1) and electromagnetic guiding (for pulse 2). 

\begin{figure}[!ht]
	\centering
	\includegraphics{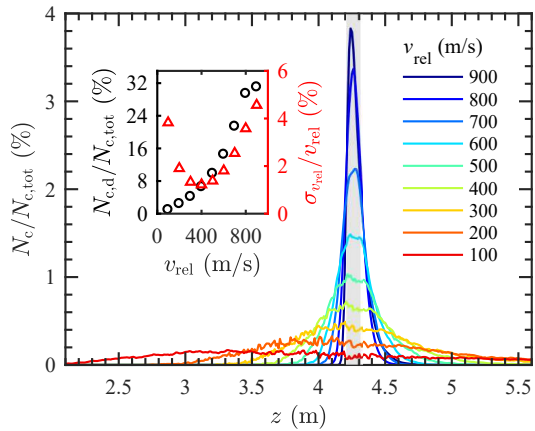}
	\caption[]{Distributions of collision numbers along the beam axis obtained from calculations at different relative beam velocities (see legend). The detector region is indicated by a gray shading. The collision numbers $N_{\mathrm{c}}$ (binned into 1 cm-long intervals) are normalized to the total number of collisions $N_{\mathrm{c,tot}}$. Inset: Detection efficiency (black circles) and fractional spread of relative velocities (red triangles) as a function of relative beam velocity, respectively.}
	\label{fig:etuning}
\end{figure}

Fig. \ref{fig:etuning} illustrates the spatial collision distributions obtained for calculations at different relative beam velocities. Here, the initial number of particles in each beam is $N_{1,0} = N_{2,0} = 1\cdot10^4$, so that $N_{\mathrm{c,tot}} = 1\cdot10^8$. A beam deflector is operated such that only the Zeeman-slowed He atoms in the $2^3$S$_1$ state, i.e., all atoms of pulse 1 with a velocity below 1000 m/s ($\approx$ 50 \% at $v_{\mathrm{rel}} \geq 100$ m/s), are admitted towards the detection region. To achieve this, all atoms outside a time window of 450 $\mu$s are deflected in the transverse direction by laser radiation resonant with the $2^3$S$_1$ -- $2^3$P$_2$ transition. A chopper wheel is not implemented (see below for details). As can be seen from Fig. \ref{fig:etuning}, the distribution of collision positions along the beam axis is strongly compressed as the relative velocity increases. This effect can be explained by the increase in interaction time with the cooling laser and the decrease in free-flight time after Zeeman slowing at higher relative velocities. The inset to the figure shows that up to one third of all collisions along the beam axis (and therefore up to 60 $\%$ of the collisions after beam deflection) are predicted to occur inside the detection region. At the same time, the fractional spread of relative velocities is found to be small (between 1--5 $\%$). The contribution of intrabeam collisions to the signal intensity on the detector is expected to be negligible, because the relative number of intrabeam collisions is decreased to below 0.1 \% at $z \geq$ 1.7 m both in the absence and in the presence of Zeeman slowing (at $v_0$ = 1000 m/s). The overall number of intrabeam collisions along the supersonic beam axis until the detection region is not significantly increased ($\leq$ 10 \%) by the Zeeman slowing process.

\subsection{\label{sec:LT}Low-collision-energy regime}
The following section is focused on the performance of the proposed merged-beam method in the regime of small relative velocities, since the corresponding collision energy range is of particular interest for the study of quantum effects in chemical reactions. More precisely, we have selected $v_{\mathrm{rel}}$ = 35 m/s, which corresponds to a collision energy of 100 mK for He($2^3$S$_1$)-H$_2$ collisions. To achieve this relative velocity, $\Delta t_{\mathrm{Zee}}$ is set such that the final velocity of pulse 1 is $v_1$ = 965 m/s after Zeeman slowing. Owing to the non-zero velocity distributions of both supersonic beams, low-energy collisions can also be observed in the absence of a Zeeman slower by letting beams 1 and 2 merge at a large distance from the source. This approach is similar to the method described by Amarasinghe et al. \cite{Amarasinghe2017}. For example, for collisions in which $v_0$ = 1000 m/s and $v_{\mathrm{rel}}$ = 35 m/s, a velocity class with $v^*_0$ = 1035 m/s must be chosen from pulse 2. Below, we will therefore also illustrate the relative efficiency of low-energy-merged-beam collisions in the presence and in the absence of the Zeeman slower. 
\begin{figure}[!ht]
	\centering
	\includegraphics{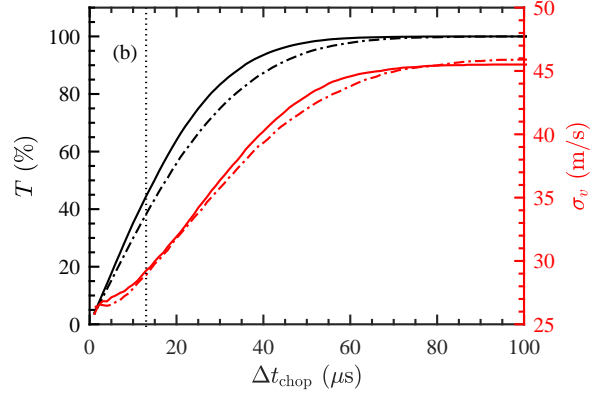}
	\caption[]{Black color: Transmission of pulse 2 through the chopper wheel as a function of chopper opening time $\Delta t_{\mathrm{chop}}$. Red color: Velocity spread of pulse 2 behind the chopper wheel as a function of $\Delta t_{\mathrm{chop}}$. Solid lines represent the results obtained from simulations in which the central part of the beam is selected (so that $v_0$ = 1000 m/s). Dash-dotted lines show results for simulations which are aimed at selecting a different part of the initial velocity distribution, for which $v^*_0$ = 1035 m/s. A dashed vertical line indicates the value of $\Delta t_{\mathrm{chop}} =$ 13 $\mu$s used for the simulation of low-energy merged-beam collisions.}
	\label{fig:choptransmission}
\end{figure}

At low collision energies, the energy resolution can be significantly improved further by cutting off the trailing edges of pulse 2 in a chopper wheel. To avoid changes in the velocity distribution of pulse 2, the chopper wheel should run at its maximum rotation speed and include one thick slit. The chopper could then be timed such that pulse 1 is fully transmitted, while pulse 2 is cut at the same point in time. At $S_{\mathrm{max}}$ = 1.3 kHz, the gas pulse duration can be reduced to 13 $\mu$s \cite{Lam2015}. Since the maximum time delay between the two beams is given by $\Delta t_{\mathrm{gp,max}} \approx 1/S_{\mathrm{max}}$ = \mbox{750 $\mu$s}, this approach is restricted to $v_{\mathrm{rel}} < 170$ m/s. Fig. \ref{fig:choptransmission} illustrates that the velocity spread of pulse 2 can be significantly reduced as the chopper opening time is decreased. For example, the velocity spread of pulse 2 is reduced by about one third when the chopper wheel is set to its minimum opening time of 13 $\mu$s. To ensure that the resulting mean velocity remains $v_0$ = 1000 m/s, the central, most intense part of pulse 2 must be transmitted (45 $\%$ of the particles, see Fig. \ref{fig:choptransmission}). As the timing of the chopper wheel is offset, a different part of the initial velocity distribution is selected, e.g., $v^*_0$ = 1035 m/s in Fig. \ref{fig:choptransmission}. As a result, the transmission through the chopper wheel is further decreased, and the velocity spread of pulse 2 is marginally altered as well (cf. Fig. \ref{fig:choptransmission}).

\begin{figure*}[!ht] 
	\centering
	\includegraphics{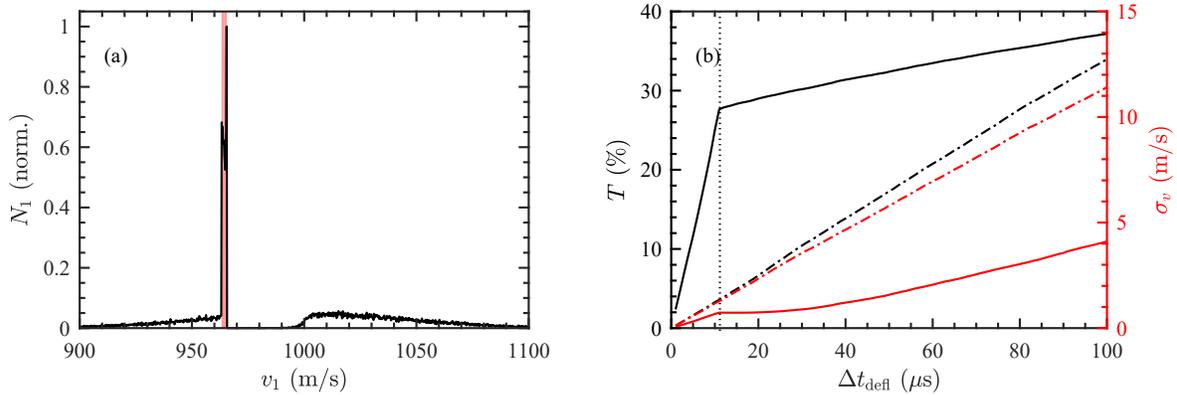}
	\caption[]{(a) Simulated velocity distribution of pulse 1 (binned into 0.2 m/s velocity intervals) at the position of the beam deflector. Here, the mean final velocity of pulse 1 after Zeeman slowing is 965 m/s. The non-deflected part of the velocity distribution (corresponding to $\Delta t_{\mathrm{defl}}$ = 11 $\mu$s) is indicated by a red shading. (b) Black color: Transmission of pulse 1 behind the deflection system as a function of time $\Delta t_{\mathrm{defl}}$, which denotes the time during which the resonant laser radiation for beam deflection is turned off. Red color: Velocity spread of pulse 2 behind the deflection system as a function of $\Delta t_{\mathrm{defl}}$. Results obtained from simulations including (excluding) the Zeeman slower are given as solid (dash-dotted) lines. A dashed vertical line indicates the value of $\Delta t_{\mathrm{defl}}$ = 11 $\mu$s used for the simulation of low-energy collisions.}
	\label{fig:defltransmission}
\end{figure*}

The transverse deflection of pulse 1 can be used as a second handle to improve the collision energy resolution at low energies. Fig. \ref{fig:defltransmission} (a) shows the velocity distribution of He atoms in the $2^3$S$_1$ state after Zeeman slowing. Since the interaction time of the atoms with the laser light for Zeeman slowing is short ($\Delta t_{\mathrm{Zee}}$ = 153 $\mu$s), both the fast and the slow part of the velocity distribution are not laser cooled. The effect of transverse deflection on the velocity spread and the transmission of pulse 1 is illustrated in Fig. \ref{fig:defltransmission} (b). Owing to phase-space compression during Zeeman slowing, the velocity spread of pulse 1 is reduced by more than a factor of 60 at $\Delta t_{\mathrm{defl}} \leq$ 21 $\mu$s. For the simulation of merged-beam collisions, $\Delta t_{\mathrm{defl}} =$ 11 $\mu$s is chosen as a compromise between a good beam transmission and a narrow velocity spread.

The velocity spread of pulse 1 can be reduced by transverse deflection even in the absence of Zeeman slowing. The results of this approach are shown as dash-dotted lines in Fig. \ref{fig:defltransmission} (b). For all values of $\Delta t_{\mathrm{defl}}$, the resulting beam transmission is lower and the velocity spread is higher than in the presence of a Zeeman slower. This comparison further illustrates the advantages of using a Zeeman slower for this kind of experiment.

In Tab. \ref{tab:compsetups}, the on-axis detection efficiencies and collision velocity resolutions of the different low-energy merged-beam approaches are listed. While the use of a Zeeman slower only (scheme ZS) yields the highest detection efficiency ($\vartheta = N_{\mathrm{c,d}}/N_{\mathrm{c,tot}} = 0.47\,\%$), the resulting spread of relative velocities is almost a factor of two higher than for other schemes ($\sigma_{v_{\mathrm{rel}}}/v_{\mathrm{rel}} = 7.3 \%$). The best compromise between a low velocity spread and a comparably high detection efficiency is the use of a Zeeman slower, a beam chopper and a deflection system (scheme ZS + BC + DS). In the absence of a Zeeman slower (scheme nZS + BC + DS), the energy resolution is further improved, but the detection efficiency is about a factor of 10 lower than in the presence of a Zeeman slower. It is also important to note that those schemes without Zeeman slowers can only be used at low collision energies and for a very narrow range of relative velocities, since they rely on the presence of different velocity classes within a supersonic beam. Therefore, the use of a Zeeman slower is indispensable for collinear merged-beam collision studies spanning a wide energy range.
\begin{table}[hbt!]
	\centering
	\caption{\label{tab:compsetups} Detection efficiency, $\vartheta = N_{\mathrm{c,d}}/N_{\mathrm{c,tot}}$, and fractional spread of relative velocities, $\sigma_{v_{\mathrm{rel}}}/v_{\mathrm{rel}}$, obtained for different low-energy, longitudinal merged-beam approaches. In the simulations, $N_{1,0} = N_{2,0} = 5\cdot10^4$. nZS = no Zeeman slower, \mbox{ZS = Zeeman} slower, BC = beam chopper ($\Delta t_{\mathrm{chop}} =$ 13 $\mu$s), DS = deflection system ($\Delta t_{\mathrm{defl}} =$ 11 $\mu$s).} 
	\begin{tabular}{lcc}
		\toprule
		& $N_{\mathrm{c,d}}/N_{\mathrm{c,tot}}$ ($\%$) & $\sigma_{v_{\mathrm{rel}}}/v_{\mathrm{rel}}$ ($\%$)\\
		\midrule
		nZS 			& 0.42 & 7.6\\
		nZS + BC 		& 0.22 & 7.9\\		
		nZS + DS		& 0.02 & 4.3\\
		nZS + BC + DS	& 0.01 & 3.0\\
		\midrule
		ZS 				& 0.47 & 7.3\\
		ZS + BC 		& 0.26 & 7.2\\		
		ZS + DS        	& 0.23 & 4.7\\
		ZS + BC + DS 	& 0.16 & 3.8\\
		\bottomrule
	\end{tabular}
\end{table}

\subsection{\label{sec:testsetup}Production of gas pulses at short time intervals}
While large time lags $\Delta t_{\mathrm{gp}}$ and $\Delta t_{\mathrm{Zee}}$ are easy to implement experimentally, it is not immediately obvious how short the time difference between the two gas pulses can be made (cf. Fig. \ref{fig:vrel} (b)), i.e., which minimum collision energy can be reached. For this reason, we have done test measurements with a room-temperature valve at different $\Delta t_{\mathrm{gp}}$ using parts of a setup described in Refs. \cite{Grzesiak2019} and \cite{Guan2019}. Here, two supersonic beams are produced by expanding room-temperature He gas from a 30 bar reservoir into the vacuum using a pulsed CRUCS valve. The first gas pulse is triggered at 200 ms intervals, while $\Delta t_{\mathrm{gp}}$ is varied in between measurements. Both He beams are excited to the metastable $2^3$S$_1$ and $2^1$S$_0$ states using an electron-seeded discharge which is attached to the front of the valve body. The discharge duration of 25 $\mu$s (full width at half maximum) is set by the valve opening time. The beam characteristics are determined using two gold-coated Faraday-cup-type detectors (FC1 and FC2) which are located in a differentially pumped vacuum chamber; FC1 is placed at a distance of 102 cm from the valve. FC1 consists of a mesh with a transmission of 73 \%, so that the beams are simultaneously measured on both detectors. The relative distance between the Faraday cups is accurately known (224.3$\,\pm\,$0.5 mm), so that the mean longitudinal velocities of the supersonic beams and the velocity widths can be determined from the measured time-of-flight (TOF) profiles at the two detectors.

\begin{figure*}[!ht] 
	\centering
	\includegraphics{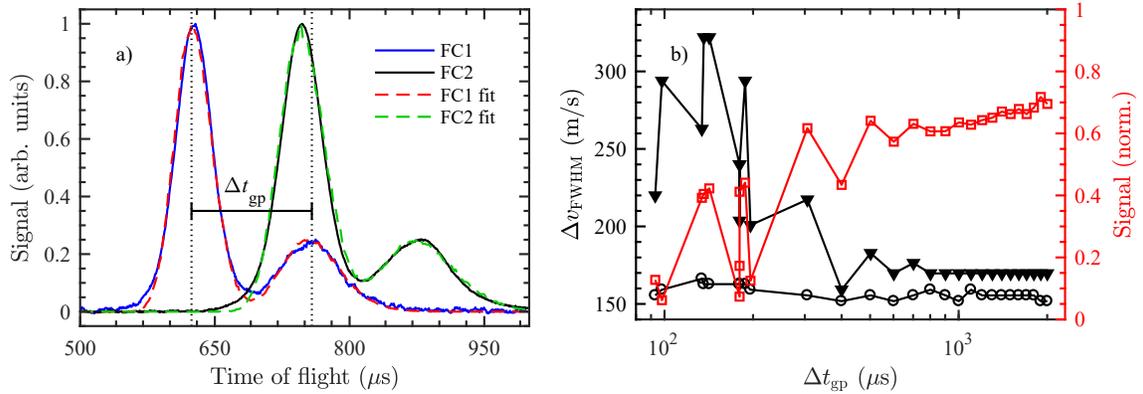}
	\caption[]{(a) Black and blue solid lines: Measured TOF profiles of two metastable He beams on two Faraday cup detectors. Red and green dashed lines: TOF distributions obtained from a numerical fit to the experimental data. The relative time delay between beams 1 and 2 is $\Delta t_{\mathrm{gp}}$ = 134 $\mu$s. (b) Typical beam characteristics as a function of time delay $\Delta t_{\mathrm{gp}}$ obtained by comparison of the experimental data with numerical simulations: longitudinal velocity spreads $\Delta v_{\mathrm{FWHM}}$ for pulse 1 (black open circles) and pulse 2 (black filled triangles), and integrated signal intensity of pulse 2 with respect to pulse 1 (red open squares). The integrated signal intensity of pulse 1 is not shown, since it remains nearly constant as $\Delta t_{\mathrm{gp}}$ is changed.}
	\label{fig:exp}
\end{figure*}
The solid lines in Fig. \ref{fig:exp} (a) show measured TOF profiles of two metastable He beams for $\Delta t_{\mathrm{gp}}$ = 134 $\mu$s, which corresponds to $v_{\mathrm{rel}}$ = 33 m/s. Even though the intensity of pulse 2 is smaller than that of pulse 1, the traces clearly show that two supersonic beams can be produced at such short time intervals. From a comparison with three-dimensional numerical trajectory calculations, we have inferred a mean velocity of 1850 m/s for both beams and longitudinal velocity spreads of $\Delta v_{\mathrm{FWHM}}$ = 160 m/s and 320 m/s for beams 1 and 2, respectively.

The typical beam characteristics have also been determined at different time delays $\Delta t_{\mathrm{gp}}$ (Fig. \ref{fig:exp} (b)). While the properties of pulse 1 and the mean velocities of both beams are nearly the same at all values of $\Delta t_{\mathrm{gp}}$, the signal intensity (longitudinal velocity spread) of pulse 2 decreases (increases) as $\Delta t_{\mathrm{gp}}$ is decreased to below 400 $\mu$s. Additional TOF measurements of atomic He with a quadrupole mass spectrometer (QMS), located in front of FC1, have allowed us to disentangle the influence of discharge excitation on the observed increase in velocity width (not shown). Both in the presence and in the absence of discharge excitation, the FWHM values of the QMS data for pulse 2 follow a similar $\Delta t_{\mathrm{gp}}$ dependence as the longitudinal velocity spread shown in Fig. \ref{fig:exp} (b). The results suggest that the increased velocity width of pulse 2 at small $\Delta t_{\mathrm{gp}}$ is likely due to beam destruction and heating caused by the high amount of gas still residing inside the source chamber when pulse 2 is produced. The QMS data also show that the FWHM values of the TOF profiles for both beams is increased when the discharge is operated compared to the case when it is turned off. This also implies that the velocity width of pulse 2 in the actual collinear merged-beam experiments will be smaller than the values given in Fig. \ref{fig:exp} (b).

\subsection{\label{sec:intrabeam}Intrabeam-collision studies}
Zeeman slowing also offers the possibility to tune the energy of intrabeam reactive collisions occuring inside of a single gas pulse. We have done additional numerical trajectory calculations to evaluate the feasibility of this alternative approach for studying low-energy collisions. In this case, only a single gas pulse (with a pulse duration $t_\mathrm{p}$) which contains both He$^*$ and the second species, is needed. In addition to that, the experimental setup is further simplified compared to the proposed collinear merged-beam approach, as it does not require a chopper wheel and a deflection system. To prevent unwanted side reactions, optical depletion can be used to depopulate the metastable $2^1$S$_0$ state of He, which is also produced during discharge excitation \cite{Hotop1969a, Guan2019}. For an initial velocity $v_0 = 1000$ m/s, the stopping distance for Zeeman slowing is $L_0$ = 213 cm. However, for the study of intrabeam collisions, the Zeeman slower can be made much shorter than that. For this reason, the overall length of the apparatus can be $<$ 150 cm.

The numerical trajectory calculations are done as described above, except that the two particle distributions ($N_1,0 = N_2,0 = 5000$) are produced at $t_0 = t_\mathrm{p}/2$ (He$^*$ atoms) and at $0 \leq t_0 \leq t_\mathrm{p}$ (second species). 

\begin{figure*}[!ht]
	\centering
	\includegraphics{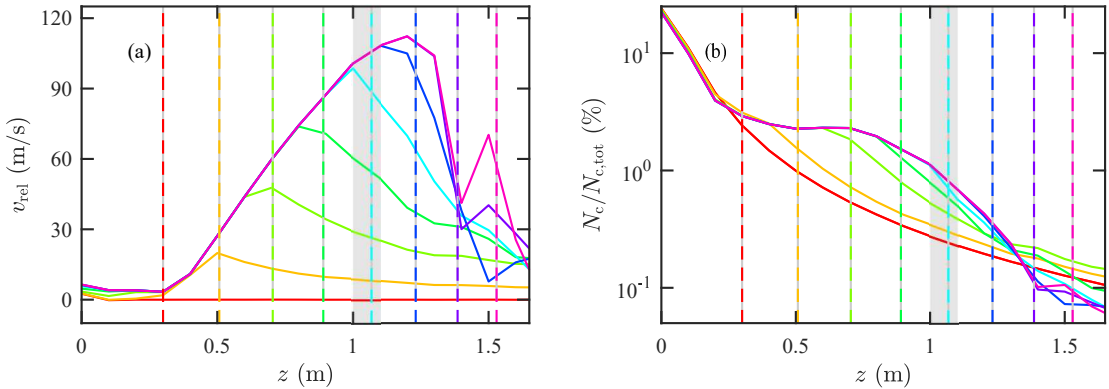}
	\caption[]{Distributions of (a) mean relative velocities and (b) normalized collision numbers along the supersonic beam axis obtained from a calculation of intrabeam collisions inside pulse 1 for different Zeeman-slowing distances (indicated by colored vertical lines). The collision numbers $N_{\mathrm{c}}$ (binned into 10 cm-long intervals) are normalized to the total number of collisions $N_{\mathrm{c,tot}}$. The detector region, which is chosen for further analysis (see Fig. \ref{fig:intrabeamzdet}), is indicated by a gray shading.}
	\label{fig:intrabeamallpos}
\end{figure*}
Fig. \ref{fig:intrabeamallpos} (a) shows the distributions of mean relative velocities along the supersonic beam axis which result from different time periods for Zeeman slowing. The times, at which the laser light for Zeeman slowing is turned off, correspond to discrete positions of the ideal particle along the supersonic beam axis $z_{\mathrm{off}}$ (indicated by colored vertical lines in Fig. \ref{fig:intrabeamallpos}). As can be seen from the figure, the relative velocity (and thus the collision energy) reaches a maximum close to these positions. The relative velocity can only be tuned over a narrow range between $0 \leq v_{\mathrm{rel}} \leq 120$ m/s. The underlying reason for this narrow tuning range is the fast separation of the Zeeman-slowed He$^*$ atoms from the non-slowed species. The maximum relative velocity does not significantly increase at higher initial velocities, i.e., $0 \leq v_{\mathrm{rel}} \leq 130$ m/s at $v_0$ = 1850 m/s. Fig. \ref{fig:intrabeamallpos} (b) illustrates that the number of collisions significantly increases owing to the Zeeman slowing process, since the He$^*$ atoms are being moved through the cloud of non-slowed species by the laser-cooling process. For example, at $z$ = 1.0 m, the number of collisions in the presence of Zeeman slowing is up to a factor of 4 higher compared to the unslowed case in which the Zeeman slower is not operated.

Since the relative velocity can only be tuned over a narrow range, it does not pay off to use a long Zeeman slower for the study of intrabeam collisions. For the further analysis, the length of the Zeeman slower is set to 1.0 m; and the detection region is in between 1.0 m $\leq z \leq$ 1.1 m. As before, the detector is placed at right angles with respect to the supersonic beam axis, so that only collisions inside the detector region are recorded. 

\begin{figure*}[!ht]
	\centering
	\includegraphics{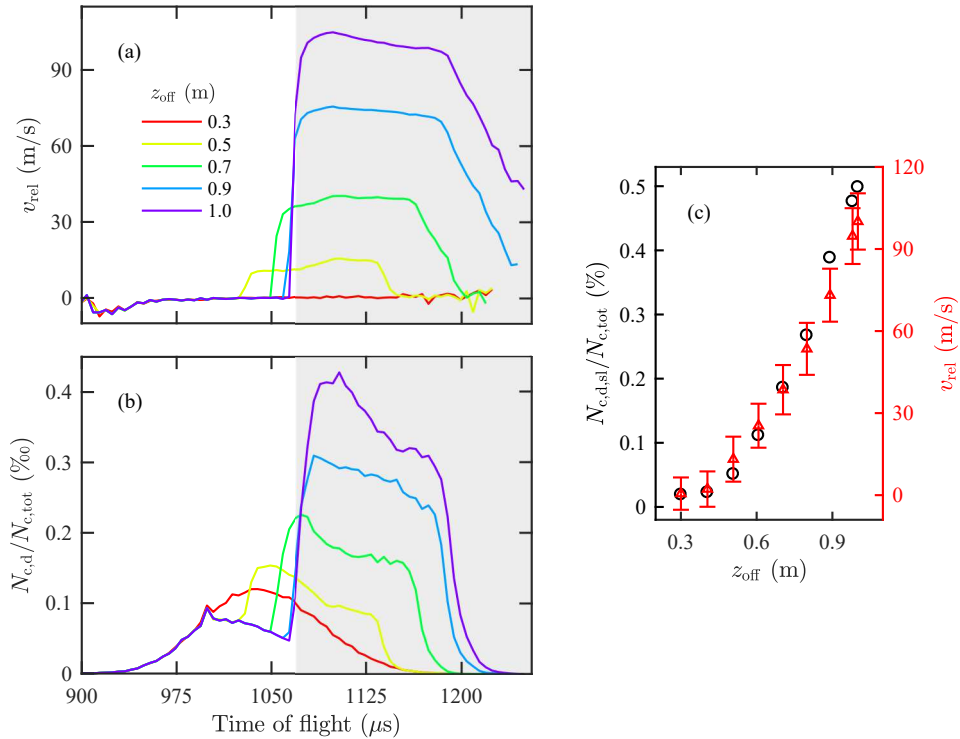}
	\caption[]{(a) Mean relative velocities and (b) normalized collision numbers inside the detector region (1.0 m $\leq z \leq$ 1.1 m) for intrabeam collisions in the presence of Zeeman slowing as a function of time of flight through the apparatus. The switch-off positions of the Zeeman slower are indicated in the legend to (a) (same color code for (a) and (b)). (c) Detection efficiency (black circles) and relative velocities (red triangles) as a function of switch-off position in a certain time interval for detection (indicated by a gray shading in (a) and (b)).}
	\label{fig:intrabeamzdet}
\end{figure*}
Figs. \ref{fig:intrabeamzdet} (a) and (b) illustrate that, for different switch-off positions $z_{\mathrm{off}}$, the mean relative velocities and normalized collision numbers inside the detector region show a considerable TOF dependency. At TOF $\leq$ 1000 $\mu$s, only non-slowed He$^*$ atoms reach the detector, so that the collision numbers are unchanged and the relative velocities are close to zero. At TOF $>$ 1000 $\mu$s, the distributions are modulated as a result of the Zeeman-slowing process. For this reason, the velocity resolution is strongly improved if only collisions with Zeeman-slowed He$^*$ atoms are sampled (e.g., gray-shaded areas in Figs. \ref{fig:intrabeamzdet} (a) and (b)). Fig. \ref{fig:intrabeamzdet} (c) clearly shows that the resulting mean relative velocity and the number of collisions are increased as $z_{\mathrm{off}}$ is increased. The velocity resolution also increases towards higher values of $z_{\mathrm{off}}$. For the selected TOF interval, the highest velocity resolution $\sigma_{v_{\mathrm{rel}}}/v_{\mathrm{rel}}$ = 10 $\%$ is achieved at $z_{\mathrm{off}}$ = 1.0 m. However, the velocity resolution can be improved even further by sampling over smaller TOF intervals.

\section{\label{sec:disc}Discussion}
The results of the numerical trajectory calculations provide information about the achievable collision-energy range and resolution, and about the detection efficiency. Below, we compare the obtained quantities with existing results from the literature, and we provide an estimate of the expected number of collision events. 

For the longitudinal merged-beam approach, the collision-energy range extends from the thermal to the quantum regime; it is only limited by the final velocity achieved by Zeeman slowing. For transverse beam merging, the relative velocity is adjusted by changing the forward velocities of the beams. If cryogenic cooling is used, the energy-tuning range is comparable for both approaches. In contrast to that, the energy-tuning range for the proposed intrabeam-collision experiment is limited to the low-energy regime, which is, however, the most interesting for the study of quantum-resonance effects \cite{Henson2012, Jankunas2015b}. 

The collision-energy spread for the proposed longitudinal merged-beam experiment, which results from the velocity width of both beams, is comparable to that obtained for transverse merged-beam techniques. For example, for He($2^3$S$_1$)-H$_2$ collisions, Shagam and co-workers \cite{Shagam2013} report an energy resolution $\Delta E_{\mathrm{c}}/E_{\mathrm{c}} =$ 6 \% (5 \%) at $E_{\mathrm{c}}=$ 1 meV (5 meV). The results of our simulations predict that $\Delta E_{\mathrm{c}}/E_{\mathrm{c}} =$ 2 \% (8 \%) at $E_{\mathrm{c}}=$ 1 meV (5 meV). The proposed method for the study of intrabeam collisions provides a lower collision-energy resolution compared to the merged-beam approaches. For instance, at $E_{\mathrm{c}}=$ 0.07 meV, $\Delta E_{\mathrm{c}}/E_{\mathrm{c}} =$ 19 \% for the intrabeam-collision experiment, whereas $\Delta E_{\mathrm{c}}/E_{\mathrm{c}} =$ 8 \% and 13 \% for the proposed merged beam approach and for the experiments by Shagam et al., respectively.

Since we are considering point-like particles, the results of our trajectory calculations do not provide a direct measure of the expected number of collision events. As a result of phase-space compression in a Zeeman slower, all merged-beam collisions are predicted to happen in a relatively small region along the beam axis, especially at relative velocities \mbox{$\geq$ 300 m/s}. However, owing to the length of the proposed setup, the transverse expansion of the two supersonic beams is much more critical than in other approaches. For example, the ratio of beam areas in the detection region is about a factor of 10 higher than in the experiment by A. Osterwalder \cite{Osterwalder2015}.

The number of collision events $N_{\mathrm{ev}}$ can be estimated using
\begin{equation}
N_{\mathrm{ev}} \approx n_{1} \, n_{2} \, \sigma V_{\mathrm{d}} \, v_{\mathrm{rel}} \, \vartheta \, t_{\mathrm{p}}, 
\label{Ncollevents}
\end{equation}
where $n_{1}$ and $n_{2}$ are the local densities of the interacting species 1 and 2 inside the detector volume $V_{\mathrm{d}}$, respectively, and $\sigma$ is the reaction cross section (e.g., $\sigma \approx$ 100 $\AA^2$ for Penning ionization). The local particle densities can be approximated using a volume dilution factor $\frac{V_{0}}{V}$ which takes into account the increase in particle volume from the source to the detector, so that $n_{i} = n_{i,0} \frac{V_{0}}{V}$ (with $i = 1,2$). The typical metastable atom densities in the source region $n_{1,0}$ are typically a factor of $10^4$ lower than for the second species \cite{Grzesiak2019}, e.g., $n_{1,0} = 10^{10}$ atoms/cm$^3$ and $n_{2,0} = 10^{14}$ particles/cm$^3$. Using the setup described in the experimental section, we estimate that $V_{\mathrm{d}} = 84$ cm$^3$ (5 cm$^3$), $\frac{V_{0}}{V} \approx 2\cdot10^{-6}$ ($1\cdot10^{-4}$) for the collinear merged-beam approach (intrabeam-collision approach). About two collision events per sequence of two pulses are thus to be expected at $E_{\mathrm{c}}=$ 4 meV for the proposed merged-beam approach. At low collision energies, this experimental approach appears to be unfeasible for reactive scattering experiments with He$^*$, e.g., $N_{\mathrm{ev}} = 7\cdot10^{-4}$ at $E_{\mathrm{c}}=$ 0.07 meV, if the beams are allowed to freely expand in the transverse direction. The study of intrabeam collisions appears to be a promising approach for studying low-energy reactive collisions of He$^*$, since $N_{\mathrm{ev}} = 1$ at $E_{\mathrm{c}}=$ 0.07 meV. The higher count rate is mainly due to the much shorter length of the experimental setup, which results in a higher local density of the collision partners. One could think of using a significantly shorter setup in order to increase the count rate for collinear merged-beam scattering, e.g., by placing the detector at the exit of the Zeeman slower and/or by using a Zeeman slower of decreased length. For example, with a He$^*$ Zeeman slower of 20 cm length, it is already possible to reach a relative velocity \mbox{$v_{\mathrm{rel}}\leq50$ m/s} (at $v_0 = 1000$ m/s). Owing to the low photon cycling rates, the laser cooling of He$^*$ requires particularly long Zeeman slowers. For other laser-coolable species, $L_0$ is typically $\leq 1$ m, so that the overall setup would also be much more compact.

\section{\label{sec:concl}Conclusions}
In this article, we have assessed -- through simulations and preliminary experiments -- the possibility of using a single supersonic beam source and a Zeeman slower for merged-beam reactive collision studies.
 
The proposed methods offer a number of benefits compared to other merged-beam approaches. Owing to the use of a single molecular beam source, the experiment can be set-up in a straightforward manner, and it is less prone to misalignment compared to other merged-beam designs. The techniques can be used for a wide range of collision systems, in which one of the species is a laser-coolable atom or molecule that can be produced inside a supersonic beam (e.g., via electron-impact excitation or photodissociation) or that can be entrained into the gas flow (e.g., via laser ablation). These processes include autoionizing collisions, ion-molecule reactions and reactions with alkali atoms. Since the Zeeman slowing process only addresses atoms in a specific electronic state, the collinear merged-beam method is both quantum-state- and species-selective. Quantum-state selection is more difficult to achieve for the intrabeam-collision approach, since several different species may be produced initially. The low number of collision events is expected to the bottleneck of the proposed methods. The study of intrabeam collisions appears to be the more promising method, since much higher count rates are expected to occur at low energies owing to the shorter overall length of the experiment. This experimental approach is also appealing owing to its technical simplicity. To significantly increase the number of collision events for both approaches, collision studies with laser-coolable species other than He$^*$ should be considered, since they do not require long Zeeman slowers. In addition to that, we suggest the transverse optical collimation of the laser-coolable species, as explained in the outline of the experiment. Pulse 2 could be confined in the transverse direction using an electrostatic or magnetic guiding field depending on whether the atom or molecule in pulse 2 has an electric or magnetic dipole moment. Since only specific quantum states are electromagnetically guided, this approach even allows for further quantum-state control of the collision system.

Instead of Zeeman slowing, also other beam deceleration techniques, such as Stark or Zeeman deceleration \cite{van2012}, could be used for merged- or intrabeam-collision experiments. These techniques allow for quantum-state selection, transverse guiding and a more versatile choice of collision partners.

\begin{acknowledgement}
	
This work is financed by the German Research Foundation (DFG) under projects DU1804/1-1 and GRK 2079. K. Dulitz acknowledges support by the Chemical Industry Fund (FCI) through a Liebig Fellowship. M. van den Beld Serrano is grateful for financial support by the Erasmus+ programme of the European Union. We thank the anonymous referees for their numerous comments and suggestions.

\end{acknowledgement}

%
%

\bibliography{mergedbeambib}

\newpage
\begin{figure*}[!ht]
	\centering
	\includegraphics{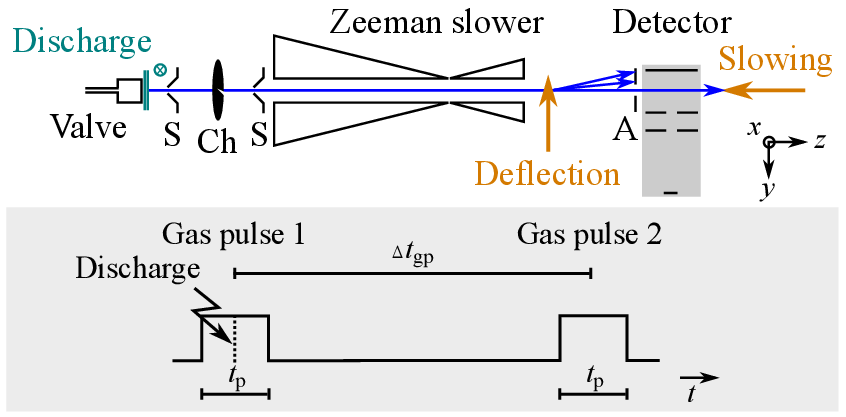}
	\caption[]{TOC Graphic.}
	\label{fig:TOC}
\end{figure*}

\end{document}